\documentclass[%
reprint,
superscriptaddress,
prl,
 amsmath,amssymb,
 aps,
]{revtex4-2}

\usepackage{lipsum}

\usepackage{graphicx}
\usepackage{dcolumn}
\usepackage{bm}

\usepackage[normalem]{ulem}
\usepackage{amsmath}
\usepackage[colorlinks=true,
					linkcolor=blue,
					urlcolor=blue,
					citecolor=blue,
					bookmarks=true,
						pdfborder={0 0 0}]{hyperref}

\begin{document}

\preprint{APS/123-QED}

\title{Social polarization promoted by sparse higher-order interactions
}

\author{Hugo P\'erez-Mart\'inez}
\affiliation{GOTHAM lab, Institute for Biocomputation and Physics of Complex Systems (BIFI), University of Zaragoza, 50018 Zaragoza (Spain).}
\affiliation{Department of Condensed Matter Physics, University of Zaragoza, 50009 Zaragoza (Spain).}

\author{Santiago Lamata-Ot\'in}
\affiliation{GOTHAM lab, Institute for Biocomputation and Physics of Complex Systems (BIFI), University of Zaragoza, 50018 Zaragoza (Spain).}
\affiliation{Department of Condensed Matter Physics, University of Zaragoza, 50009 Zaragoza (Spain).}

\author{Federico Malizia}
\affiliation{Network Science Institute, Northeastern University London, London E1W 1LP, United Kingdom}

\author{Luis Mario Flor\'ia}
\affiliation{GOTHAM lab, Institute for Biocomputation and Physics of Complex Systems (BIFI), University of Zaragoza, 50018 Zaragoza (Spain).}
\affiliation{Department of Condensed Matter Physics, University of Zaragoza, 50009 Zaragoza (Spain).}

\author{Jes\'us G\'omez-Garde\~nes}
\thanks{gardenes@unizar.es}
\affiliation{GOTHAM lab, Institute for Biocomputation and Physics of Complex Systems (BIFI), University of Zaragoza, 50018 Zaragoza (Spain).}
\affiliation{Department of Condensed Matter Physics, University of Zaragoza, 50009 Zaragoza (Spain).}
\affiliation{Center for Computational Social Science, University of Kobe, 657-8501 Kobe (Japan).} 

\author{David Soriano-Pa\~nos}
\thanks{sorianopanos@gmail.com}
\affiliation{GOTHAM lab, Institute for Biocomputation and Physics of Complex Systems (BIFI), University of Zaragoza, 50018 Zaragoza (Spain).}
\affiliation{Departament d’Enginyer\'ia Inform\`atica i Matem\`atiques, Universitat Rovira i Virgili, 43007 Tarragona, Spain.}

\date{\today}

\begin{abstract}
Many social interactions are group-based, yet their role in social polarization remains largely unexplored. 
To bridge this gap here we introduce a higher-order framework that takes into account both group interactions and homophily.
We find that group interactions can strongly enhance polarization in sparse systems by limiting agents’ exposure to dissenting views. Conversely, they can suppress polarization in fully connected societies, an effect that intensifies as the group size increases.
Our results highlight that polarization depends not only on the homophily strength but also on the structure and microscopic arrangement of group interactions.
\end{abstract}

\maketitle

\section{Introduction}

Understanding how human interactions in large-scale societies generate consensus, cooperation, political identities, or persistent divisions remains a central challenge in modern social science~\cite{boyd2007mathematical,axelrod}. 
During the last two decades, the statistical physics community has devoted substantial effort to explaining how macroscopic collective behavior can emerge from simple microscopic interaction rules~\cite{RevModPhys.81.591,JUSUP20221}.  
Among these phenomena, the emergence of \textit{polarization} has become a paradigmatic problem, whose relevance continues to grow due to the increasing availability of data from digital media \cite{kubin2021role,cinelli2020covid,falkenberg2022growing}.
\medskip

Mathematical models of opinion dynamics capable of reproducing polarization have been proposed in numerous studies~\cite{sasahara2021social,tokita2021polarized,meng2018opinion,santos2021link,baumann2020modeling,baumann2021emergence,perez2023polarized}.
These models typically assume that: (i) agents gradually discount their current opinion, (ii) they tend to align with their neighbors, and (iii) like-minded peers exert a stronger influence.
The latter captures two well-established observations: \textit{homophily}, i.e., the tendency to interact with and imitate similar others~\cite{mcpherson2001birds,bessi2016homophily}, and \textit{group polarization}, whereby intragroup agreement drives individuals towards more extreme positions~\cite{sunstein2007group}.
\medskip

As with other socio-physical models \cite{PhysRevE.74.056108,PhysRevE.77.041121,PhysRevLett.100.108702,GOMEZGARDENES2008296,PhysRevE.67.026120,PhysRevE.81.056105,doi:10.1073/pnas.0508201103}, most mathematical formulations represent the backbone of human interactions as a complex network \cite{latora2017complex,newman2018networks}. Yet, in real-world settings, opinions are rarely shaped through isolated dyadic interactions, since they are formed, reinforced, or challenged within actual groups of more than two individuals. From informal discussions to deliberative assemblies, those higher-order (i.e., group-based) interactions  \cite{battiston2020networks,bick2023higher,majhi2022dynamics,iacopini2019simplicial,lamata2024integrating} represent the native environment of opinion formation. 
\medskip

Although recent studies have incorporated higher-order influence into consensus dynamics, either through diffusion-like processes~\cite{neuhauser2020multibody,neuhauser2021consensus,sahasrabuddhe2021modelling} or by generalizing bounded confidence models~\cite{deffuant2000mixing,schawe2022higher,hickok2022bounded}, the role of group structure in driving polarization remains largely unexplored. In particular, existing approaches fail to capture the sharply bimodal opinion distributions observed in empirical data~\cite{fiorina2008political,ANES2017}.
\medskip

\begin{figure*}[t!]
    \centering
    \includegraphics[width=0.81\linewidth]{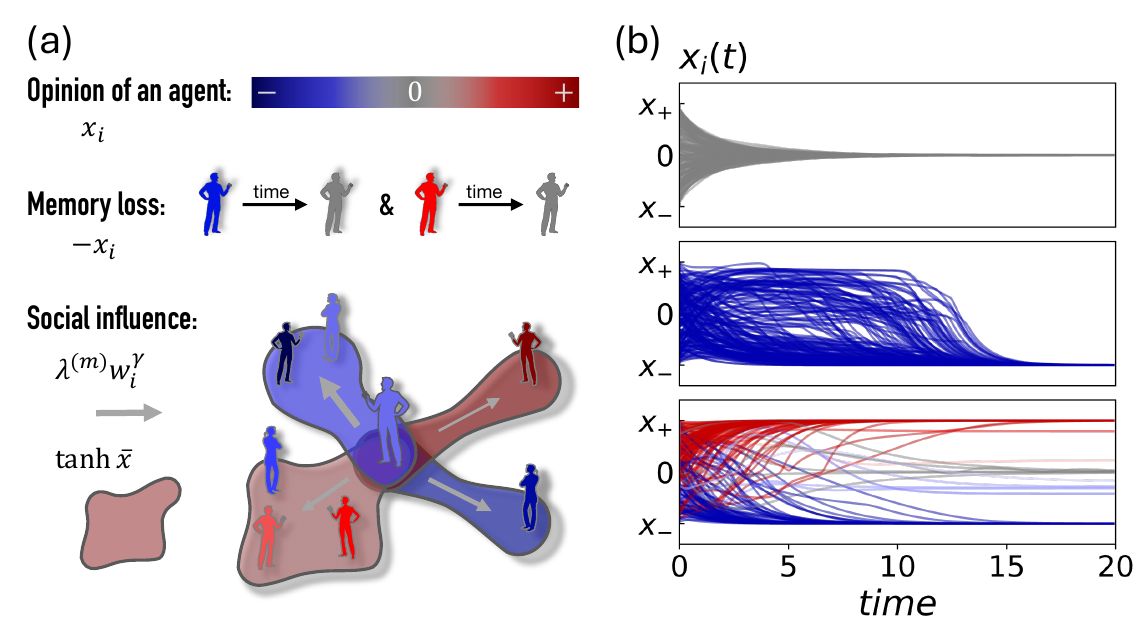}
    \caption{\textbf{Opinion dynamics with higher-order interactions} (a) Schematic representation of the model: the memory loss term pulls agents toward a neutral opinion, while the social influence term pulls agents toward the opinions of their acquaintances. (b) Temporal evolutions of agents in the representative configurations: neutral consensus (top, $\lambda^{(1)} = \lambda^{(2)} = 0.4$, $\beta = 0.7$), radicalization (center, $\lambda^{(1)} = \lambda^{(2)} = 10$, $\beta = 0.2$)
and polarization (bottom, $\lambda^{(1)} = \lambda^{(2)} = 10$, $\beta = 1.5$. All results are for a Random Simplicial Complex (RSC) of $N=1899$ nodes, $\langle k^{(1)} \rangle=10$ and $\langle k^{(2)}\rangle=3$. Initial opinions are randomly chosen on the interval $[x_-,x_+]=[-20, 20]$ (except for the top panel, where $[x_-,x_+]=[-1,1]$), and the final opinion of the agents determines the color code.}
    \label{fig:Fig0}
    \vspace{-1.5em}
\end{figure*}

In this article, we introduce a dynamical framework for opinion formation that extends empirically grounded mechanisms to explicitly incorporate group interactions, encoded via higher-order networks. We find that group interactions promote radical consensus only when they dominate the dynamics. In contrast, when pairwise interactions prevail, sparse higher-order interactions enhance polarization by reinforcing local agreement within groups. Furthermore, we show through analytical results that the impact of group interactions intensifies with increasing group size. 
Our findings show that polarization depends not only on the homophily strength but also on the structural arrangement of social interactions, providing new insights into how group dynamics shape collective opinion formation.

\section{Results}

\subsection{Opinion dynamics with higher-order interactions}

We model the structure of social interactions by means of a hypergraph $\mathcal{H}$ formulation. This type of structure is defined as a pair $\mathcal{H} = (\mathcal{N}, \mathcal{E})$, where $\mathcal{N}$ is the set of $N$ nodes (here the agents), and $\mathcal{E}$ is a collection of subsets of $\mathcal{N}$ called \textit{hyperedges}. Each hyperedge $\gamma \in \mathcal{E}$ represents a group of agents that interact simultaneously. The \textit{order} of a hyperedge is given by the number of agents it connects minus one; that is, a hyperedge $\gamma$ of size $m+1$ is said to be of order $m$. This way, $1$-hyperedges correspond to pairwise interactions, $2$-hyperedges to triplets, and so on, up to the maximum order of interaction $M$. The set $\Gamma^{(m)}_i$ is the collection of all the hyperedges of order $m$ to which node $i$ belongs and its cardinality corresponds to the generalized degree of order $m$ of node $i$, defined as $k_{i}^{(m)}$.
\medskip

To investigate how higher-order interactions shape collective opinion dynamics, we consider a framework in which each agent $i$ holds a continuous opinion variable $x_i \in \mathbb{R}$, whose sign $\sigma(x_i)\equiv \sigma_i$ represents the qualitative position of the agent (in favor or against). In its turn, the module $|x_i|$ captures the strength of her belief. 
\medskip

The opinion of each agent, i.e., the value of $x_i$, evolves in time due to their associated group-based interactions according to:
\begin{equation}
    \dot{x}_i = -x_i(t) + \sum_{m=1}^M \lambda^{(m)}\left[\sum_{\gamma \in \Gamma_i^{(m)}} w^{\gamma}_i(t) \tanh\left(\sum_{\substack{j\in \gamma \\ j \neq i}} \frac{x_j(t)}{m}\right) \right] \;.\label{eq:der_x} 
\end{equation}
This set of coupled differential equations generalizes the model introduced in Ref. \cite{perez2023polarized} to include higher-order interactions. The first term describes memory loss, while the second term pulls agents in the direction of their contacts' opinions (specially those more radical ones). Additionally, $\lambda^{(m)}$ represents the \textit{social interaction strength} of hyperedges of order $m$ over the agent's opinion, and $w^{\gamma}_i$ represents the importance that the agent gives to hyperedge $\gamma$, and thus, the influence that the group exerts over the agent (see Fig. \ref{fig:Fig0}.a for a schematic representation of the model). Each weight $w^{\gamma}_i$ evolves in time since it depends on the opinions of the agents in hyperedge $\gamma$, being its precise form: 
\begin{equation}
    w^{\gamma}_i = \frac{\left(\sum_{j\in \gamma}|x_i-x_j| + \epsilon^{(m)}\right)^{-\beta}}{\sum_{\xi \in \Gamma_i^{(m)}}\left(\sum_{l\in \xi}|x_i-x_l| + \epsilon^{(m)}\right)^{-\beta}} \, ,
    \label{eq:weights}
\end{equation}
where $\beta$ is the homophily parameter, and $\epsilon^{(m)}$ is a small regularizing constant (set to $\epsilon^{(m)} = 0.002\lambda^{(m)}$) introduced to avoid divergences for nearly identical opinions. This definition ensures that groups with agents having opinions more aligned with that of agent~$i$ exert a stronger influence, thus capturing the effect of homophily at the level of hyperedges. 
\medskip

Altogether, the dynamics of the system gives rise to three qualitatively distinct collective states: a \textit{neutral consensus}, where all agents converge to a moderate opinion; a \textit{radical consensus}, characterized by unanimous but extreme views; and a \textit{polarized state}, in which the population splits into opposing opinion blocks.
Figure~\ref{fig:Fig0}.b displays representative opinion trajectories for $M=2$, corresponding to the three collective states described above. The dynamics were simulated on a Random Simplicial Complex (RSC) consisting of $N = 1899$ nodes, with average degree $\langle k^{(1)} \rangle = 10$ for pairwise interactions and a sparser higher-order connectivity characterized by $\langle k^{(2)} \rangle = 3$, in which the subjacent $2-$hyperedge network is connected. Further information on how these configurations are built can be found in the Methods section and in Ref.~\cite{iacopini2019simplicial}.
\medskip

Notably, as derived in the Methods section, reaching neutral consensus is only possible when the total interaction strength remains below the critical threshold,
\begin{eqnarray}
\label{eq:threshold}
\sum_{m=1}^M \lambda^{(m)} < 1\;,
\end{eqnarray}
so that the first (dissipative) term in Eq.~(\ref{eq:der_x}) prevails and most of opinions concentrate around $0$. Remarkably, despite the non-linear nature of the dynamics, higher-order interactions contribute linearly across orders, with each $\lambda^{(m)}$ playing a direct role in driving the system away from neutral consensus. If the combined strength satisfies Eq.~\eqref{eq:threshold}, the system reaches polarized or radicalized configurations.
\medskip

\subsection{Sparse higher-order interactions promote polarization}

Now we move beyond the neutral consensus scenario by investigating how the interplay between social interaction strength, group structure, and homophily governs the emergence of either polarization or radicalization. To this aim, we fix $M=2$, a large total interaction strength, $\lambda^{(1)} + \lambda^{(2)} = 20$, and we define $\delta = \lambda^{(2)}/20$ to interpolate between purely pairwise ($\delta = 0$) and purely higher-order ($\delta = 1$) regimes. As a proxy for polarization, we measure the fraction of configurations in which the standard deviation of opinions exceeds the absolute value of the mean, i.e., $\sigma > |\bar{x}|$, where $\bar{x} = \sum_i x_i/N$ and $\sigma^2 = \sum_i (x_i - \bar{x})^2/N$. 
\medskip

\begin{figure}[t!]
    \centering
    \includegraphics[width=1\linewidth]{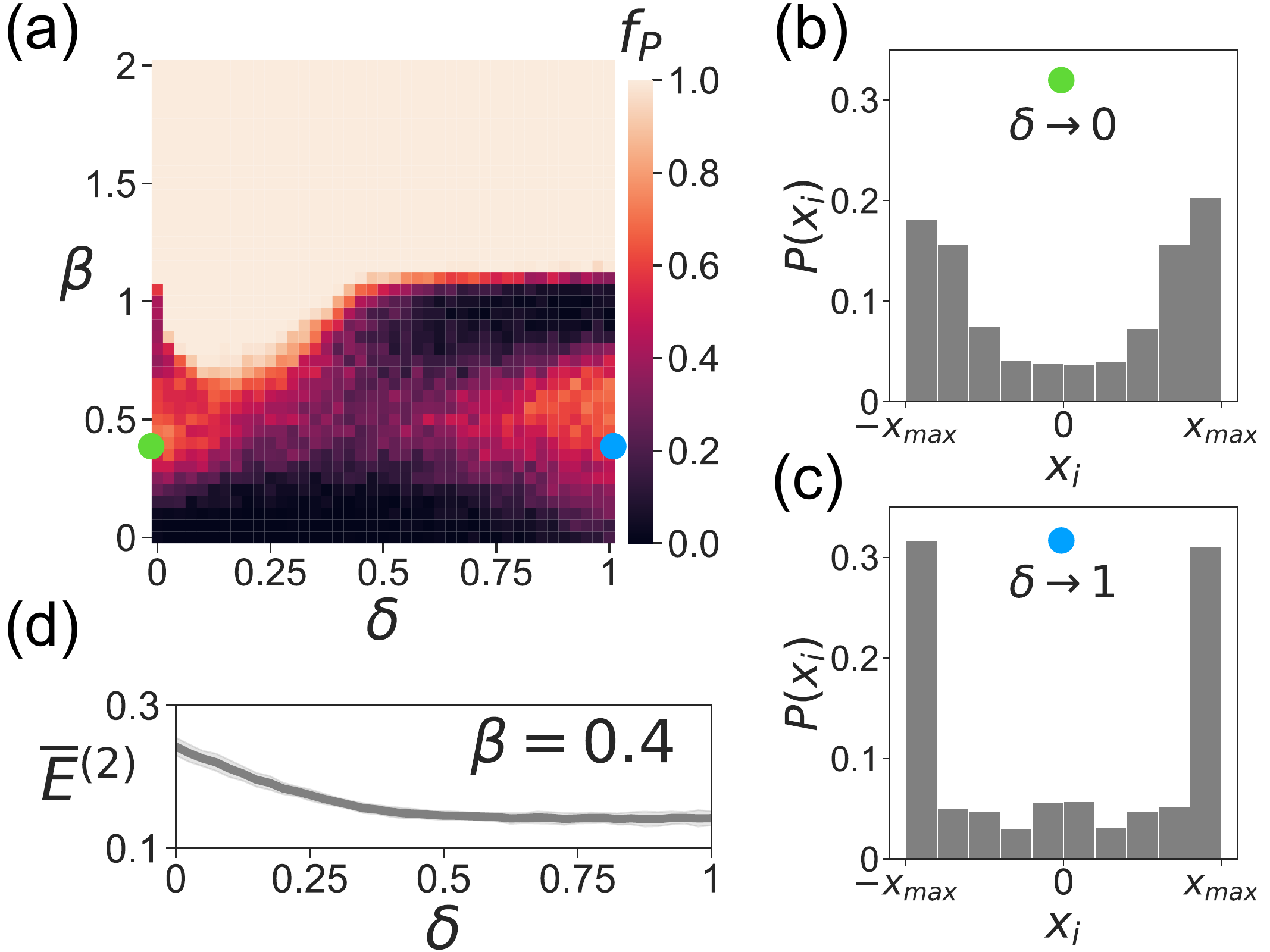}
    \caption{\textbf{Sparse higher-order interactions promote polarization.} (a) Fraction of polarized configurations $f_P$ obtained from 100 independent realizations for each combination of parameters $(\delta,\beta)$ (color code). (b) and (d) Average opinion histograms obtained from 100 polarized configurations for $\delta=0$ and $\delta=1$ given $\beta=0.4$. (c) Average exposure $\bar{{E}}^{(2)}$ as a function of $\delta$ for $\beta=0.4$, obtained from 100 polarized configurations for each parameter combination.
    All results are for an RSC of $N=1899$ nodes, $\langle k^{(1)} \rangle=10$ and $\langle k^{(2)}\rangle=3$. Initial opinions are randomly chosen on the interval $[-20, 20]$.}
    \label{fig:networked}
\vspace{-1.5em}
\end{figure}

Figure~\ref{fig:networked}.a shows the fraction of polarized configurations as a function of the higher-order weight $\delta$ and the homophily parameter $\beta$. For each parameter set, we perform 100 independent runs with uniformly distributed random initial opinions (see Methods for the numerical integration details). For $\delta = 0$ (pairwise interactions only), we observe a clear transition from mixed states (polarization and radical consensus) to fully polarized configurations at $\beta_c =1$. Interestingly, polarization is already prominent around $\beta \simeq 0.4$, a regime in which the low value of homophily is compensated by the limited neighborhood of the agents imposed by the underlying structure, which facilitates the emergence of opposing radical clusters. These findings align with those reported in Ref.~\cite{perez2023polarized}, which focuses on strict pairwise interactions. As $\delta$ increases, the system becomes increasingly susceptible to polarization, noticeable from the decrease in the critical value of $\beta$ required for its onset. However, this behavior is softened for $\delta \gtrsim 0.2$: the threshold $\beta$ for full polarization starts to rise, and for $\delta \gtrsim 0.5$ the system recovers the usual transition at $\beta_c=1$. 
\medskip

\begin{figure*}[t]
    \centering
    \includegraphics[width=0.775\linewidth]{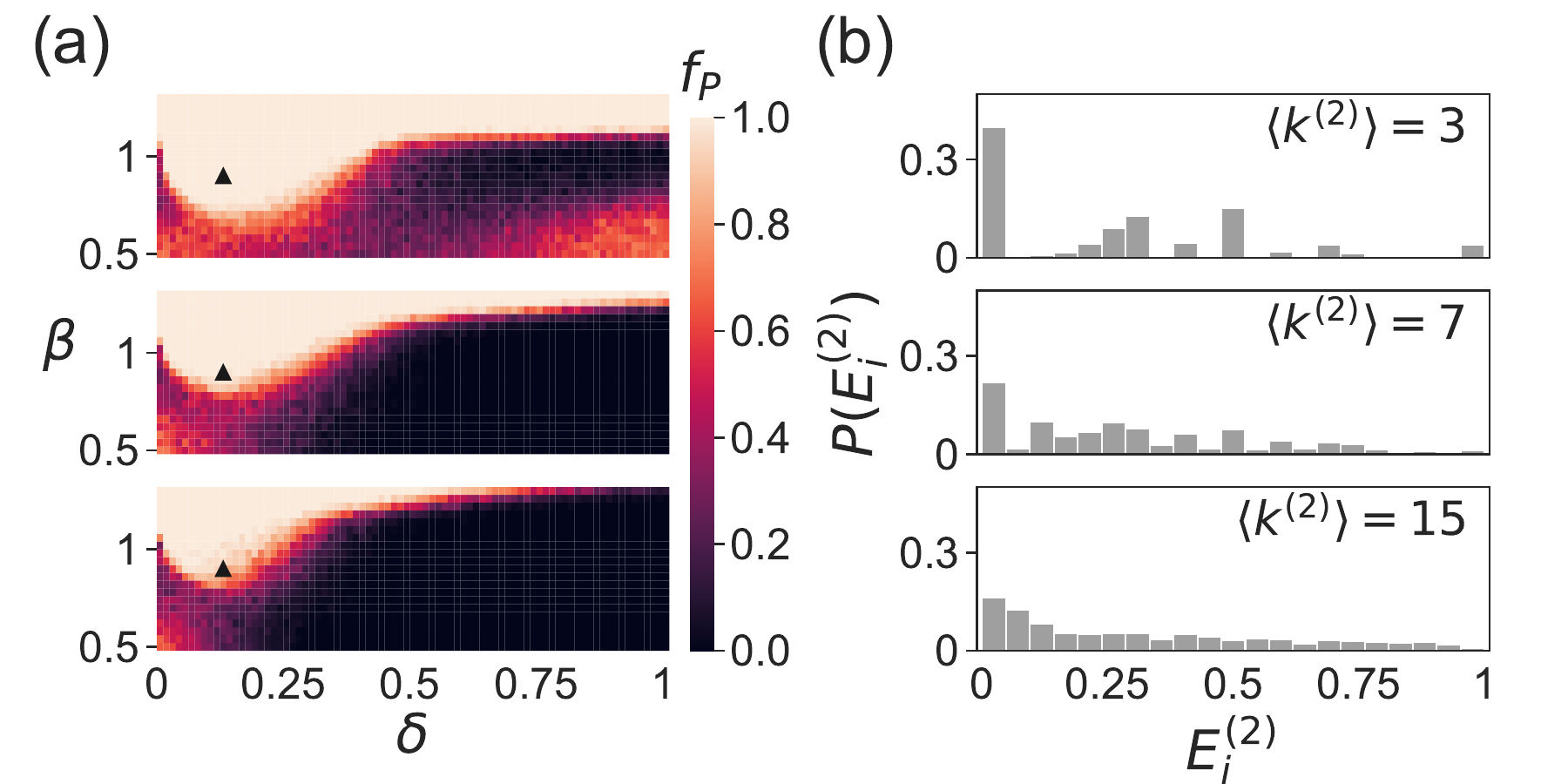}
    \caption{\textbf{Dense group interactions suppress polarization.} (a) Fraction of polarized configurations obtained from 100 independent realizations for each combination of parameters $(\delta,\beta)$ (color code), for structures with different $\langle k^{(2)}\rangle$, $\langle k^{(1)}\rangle=10$, and maximum inter-order hyperedge overlap. Networks are comprised of $N=2000$ nodes, except the graph of $\langle k^{(2)}\rangle=3$, which has $N=1899$. (b) Distribution of exposures $E_i^{(2)}$ for 100 polarized configurations generated with parameters $\beta=0.9$ and $\delta=0.125$ (pinpointed with a black triangle in panel a). Initial opinions are randomly chosen on the interval $[-20, 20]$.}
    \label{fig:exposures}
\end{figure*}

These results uncover a nontrivial role of higher-order interactions in shaping collective dynamics. When pairwise interactions dominate, a sparse group structure amplifies polarization by reinforcing local agreement within sparsely connected clusters. In contrast, when higher-order interactions dominate, their effect suppresses this mechanism, apparently recovering the dynamical effects observed in purely pairwise-only scenarios. To illustrate the qualitative differences between the interaction regimes, Fig.~\ref{fig:networked}.b and Fig.~\ref{fig:networked}.c show the distribution of the polarized configurations at fixed homophily $\beta = 0.4$, comparing the pairwise limit ($\delta \to 0$) to the regime dominated by higher-order interactions ($\delta \to 1$). In the latter case, opinions concentrate sharply around the extremal values, reflecting a population split into cohesive groups holding opposite-sign views.
\medskip

To better understand the mechanism behind this phenomenon, we define the \textit{exposure} of an agent $i$ to disagreement within its higher-order neighborhood as:
\begin{eqnarray}
    E_i^{(2)} = \frac{1}{|\Gamma_i^{(2)}|}\sum_{\gamma\in \Gamma_i^{(2)}}(1-\delta_{\sigma_i,\sigma_{\gamma}})\;.
    \label{eq:exposure}
\end{eqnarray}
In this expression $\Gamma_i^{(2)}$ is the set of 2-hyperedges involving agent $i$, and $\sigma_\gamma$ denotes the sign of the average opinion within group $\gamma$, excluding $i$. The quantity $E_i^{(2)}$ measures the fraction of \textit{cross-cutting} higher-order interactions, namely those where the position of the group opposes that of the agent. Equipped with this metric, we can capture the local exposure to dissent, which shapes polarization at the microscopic level. Note that a triangle is not classified as cross-cutting if the average group opinion aligns with $\sigma_i$, even when one member disagrees; this occurs when the third agent is sufficiently radicalized to compensate.
\medskip

Figure~\ref{fig:networked}.d shows the average exposure $\bar{E}^{(2)}=N^{-1}\sum_{i=1}^NE_i^{(2)}$ as a function of $\delta$ for fixed $\beta = 0.4$. In the $\delta \to 0$ regime, in which the pairwise interactions prevail, agents frequently encounter opposing opinions, leading them to intermediate positions and producing broad, mixed opinion distributions [Fig.~\ref{fig:networked}.b]. These opposing neighbors may also appear within 2-hyperedges, but their influence is minimal due to the weak higher-order coupling. In contrast, when $\delta \to 1$, group influence dominates. Even with a disagreeing neighbor present, the collective opinion of the group typically reinforces rather than challenges an agent's existing beliefs. This reflects the growing internal homogeneity of 2-hyperedges, which limits exposure to dissent and facilitates the emergence of polarized states associated with sharper opinion distributions [Fig.~\ref{fig:networked}.c].
\medskip

\subsection{Dense higher-order interactions suppress polarization}

Having established that sparse higher-order interactions promote social polarization, we now investigate the role of increasing the net structural prominence of group interactions. To do so, we consider three different higher-order networks with $\langle k^{(1)} \rangle = 10$ and different connectivities of group interactions $\langle k^{(2)} \rangle = \{ 3,7,15\}$.  Additionally, to isolate the effect of differing levels of downward closure (i.e., the tendency of higher-order hyperedges to contain lower-order ones among their subsets), we construct hypergraphs that exhibit the maximum possible inter-order hyperedge overlap~\cite{lamata2025hyperedge} for each chosen pair of macroscopic connectivities $(\langle k^{(1)} \rangle, \langle k^{(2)}\rangle)$ (see Methods for the crafting details).
\medskip

In Fig.~\ref{fig:exposures}.a, we show the fraction of polarized configurations as a function of $\delta$ and the homophily parameter $\beta$, for the three different connectivities. The results reveal that polarization is strongly suppressed as agents' structural reach increases (larger connectivity values). In particular, polarized states at intermediate values of $\beta$ disappear (around $\beta\approx0.5$), confirming that they appeared due to sparse higher-order interactions.
Moreover, the size of the polarization bump found under weak higher-order influence also shrinks as the $2-$hyperedges degree increases. 
\medskip

To explain this, Fig.~\ref{fig:exposures}.b shows exposure histograms from $100$ independent simulations inside the bump for multiple network structures. It is clear that some agents retain cross-cutting interactions ($E_i^{(2)} \neq 0$), while others are embedded in completely homogeneous environments ($E_i^{(2)} = 0$). As $\langle k^{(2)} \rangle$ increases, the fraction of agents with zero exposure decreases, as it becomes increasingly probable to find at least one cross-cutting triangle. Still, even for $\langle k^{(2)} \rangle =15$, $E_i^{(2)}=0$ remains the most frequent case due to the homophily mechanism. 
\medskip

Cross-cutting group interactions hinder polarization, as their average opinion tends to be closer to the agent’s own and more persuasive, thus increasing the likelihood of switching sides.
When higher-order influence is weak ($\delta\to0$), these effects are insufficient to reverse opinions, but they widen opinion distributions (see Fig. \ref{fig:networked}.b), which greatly enhances polarization, as we explained above. 
In contrast, when higher-order interactions are dense and become dominant ($\delta\to1$ and large connectivity), their full effect emerges: agents radicalize more, depleting intermediate positions, and polarization for intermediate values of $\beta$ fades (see Fig. \ref{fig:exposures}.a). This depolarizing effect intensifies as more agents are exposed to cross-cutting triangles. Strikingly, in this dominance regime, the critical value of homophily required for the emergence of polarized states shifts toward larger values. Additionally, in the SM we investigate how the microscopic organization of group interactions---specifically the hyperedge overlap~\cite{lamata2025hyperedge, malizia2025hyperedge}---influences the collective behaviour of the system.
\medskip

\subsection{Analytical derivation of the polarization threshold}

To quantify the shift on the polarization threshold when higher-order interactions are dense and dominant, we analyze the stability of fully polarized states in fully connected hypergraphs, where each agent participates in all possible $m$-hyperedges up to order $M$. We consider a configuration in which $N_+$ agents hold a positive opinion $x_+$ and $N_- = N - N_+$ hold a negative opinion $x_-$.
Through a linear stability analysis (detailed in Methods and Supplementary Eqs. (S.6)-(S.23)), we reveal that such polarized states are stable provided the homophily parameter $\beta$ exceeds a critical threshold $\beta_c$, determined implicitly by
\begin{equation}
\beta_c \sum_{m=1}^M \lambda^{(m)} f(m, \lambda^{(m)}, \beta_c) = 1,
\label{eq:condition}
\end{equation}
where $f(m, \lambda^{(m)}, \beta_c)$ is detailed in the Methods section.
\medskip

\begin{figure}[t!]
    \centering
    \includegraphics[width=\linewidth]{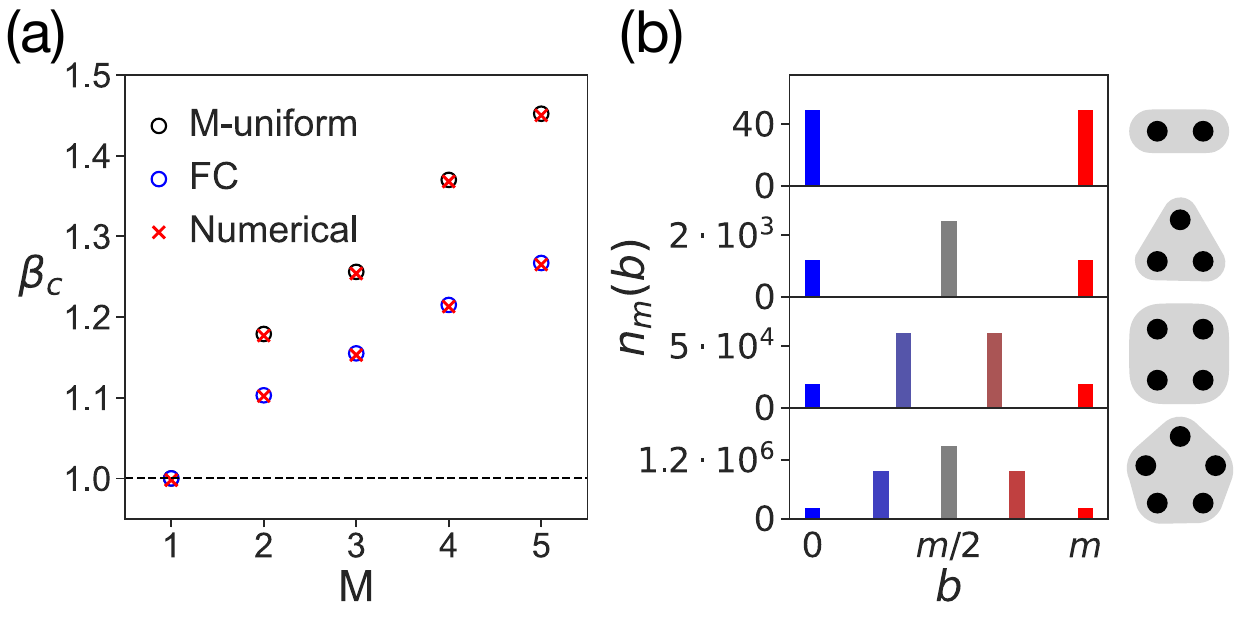}
    \caption{\textbf{Analytical derivation of the polarization threshold.} (a) Stability thresholds $\beta_c$ for a polarized configuration computed from Eq.~(\ref{eq:stab}) considering $M$-uniform hypergraphs (black circles) and fully-connected hypergraphs up to order $M$ (blue circles), as a function of the highest order present in the network. Results are shown together with the numerical values obtained by running the system of $N$ equations for validation purposes (red crosses). Throughout the figure, we consider networks comprised of $N=100$ agents. (b) Number of groups $n_m(b)$ to which an agent is exposed comprised by $b$ neighbors of opinion $x_-$ and $m-b$ neighbors of opinion $x_+$, as a function of $b$. Multiple orders of interaction are shown, from $m=1$ (top) to $m=4$ (bottom).}
    \label{fig:FC}
    \vspace{-1.5em}
\end{figure}

In Fig~\ref{fig:FC}.a we show the behavior of $\beta_c$ as a function of $M$ for two scenarios. First, we consider the case of fully connected hypergraphs with interactions up to order $M$ (with $\lambda^{(m)} = 20/M$), and secondly, $M$-uniform hypergraphs containing only order-$M$ interactions. In both cases, $\beta_c$ increases with $M$, indicating that polarization becomes progressively harder to sustain with increasing group size. Our analytical predictions are in excellent agreement with numerical integration of the full dynamics for $N = 100$ agents.
\medskip

As hinted before, the mechanism behind this effect lies on the existence of cross-cutting hyperedges that also include agreeing agents, making the group more influential while still acting as a depolarizing force, thus requiring higher values of homophily, $\beta$, to compensate.As we saw before, in the case of networked populations these heterogeneous groups could be absent for a non-negligible set of agents (those with null exposure, see Fig. \ref{fig:exposures}.a), but in the case of fully connected hypergraphs or order $m$, for a given agent holding opinion $x_+$, there are
\begin{align}
\label{eq:n_b}
    n_m(b) = {{N_+}\choose{m-b}} {{N_-}\choose{b}}
\end{align}
such groups with $b$ neighbors with opposite opinion ($x_-$), and $m-b$ with the same opinion ($x_+$).
\medskip

In Fig.~\ref{fig:FC}.b we represent $n_m(b)$ as a function of the average opinion of these groups based on the order $m$. For pairwise interactions ($m = 1$), we find that the neighbouring agents can exhibit only purely positive/negative opinions. However, as group size increases, a rich repertoire of mixed compositions emerges. In larger groups, agents often share cross-cutting $m$-hyperedges with some like-minded individuals, rather than being completely isolated. For example, a positive-opinion node in a $3$-hyperedge might encounter two negative agents and one other positive agent. While the group average remains negative, it is much closer to zero than if all three other members held negative views. In other words, larger groups that produce more moderate collective opinions include agreeing neighbors. As a result, they are more effective at pulling agents toward the center and the opposite view, resulting in radicalized societies.
\medskip

\section{Discussion}

In this article, we have introduced a dynamical framework for opinion formation that explicitly incorporates homophily and group interactions through higher-order networks. Our results reveal a fundamental asymmetry in how group structure shapes collective opinion dynamics. While sparse group interactions amplify polarization, with agents encountering limited dissenting views,  dense group interactions suppress polarization by exposing agents to more moderate collective opinions through compositional diversity. Moreover, the larger the group size, the more prominent those effects are.
\medskip

Unlike previous models, in which only pairwise interactions were considered, here the existence of higher-order interactions provides a mechanism for the emergence and loss of polarization that can be understood through the exposure of agents to opposite opinions through groups. In this respect, networked populations with sparse higher-order interactions are prone to the presence of homogeneous groups in which everyone agrees without any dissent, thus becoming a net polarizing force. However, when group interactions are more prominent in connecting the network, either because of being dense or because having low hyperedge overlap~\cite{malizia2025hyperedge,lamata2025hyperedge}, cross-cutting groups become increasingly probable while usually containing agreeing individuals at the same time. This increases their influence and drives their constituents towards the majority, encouraging agreement and reproducing the phenomenon of group polarization~\cite{sunstein2007group}.
\medskip

Indeed, 
our results are 
consistent with recent findings pinpointing that higher-order interactions favor consensus, as being exposed to intermediate opinions greatly favors agreement within a group \cite{sahasrabuddhe2021modelling,hickok2022bounded,schawe2022higher}. However, as we have discussed, this effect is contingent upon certain network characteristics, that cannot be overlooked. Our framework thus provides a foundation for the development of more advanced frameworks that capture key structural features of higher-order networks, such as their temporal evolution \cite{iacopini2024temporal,gallo2024higher,arregui2024patterns}, and how they shape the emergence of polarization in real-world systems. 
\medskip

\section{Methods}

\subsection{Crafting of higher-order structures}

\subsubsection{Random Simplicial Complex (RSC)}
In our manuscript, we considered Random Simplicial Complexes (RSCs) of order $M=2$ generated following the model introduced by Iacopini et al.~\cite{iacopini2019simplicial}. Given a set of \( N \) nodes, each pair of nodes \( (i, j) \) is connected with probability \( p^{(1)} \), resulting in an Erdős-Rényi graph with average 1-degree \( \langle k^{(1)} \rangle = (N - 1)p^{(1)} \). Moreover, each triple \( (i, j, k) \) is filled with a 2-simplex with probability \( p^{(2)} \), regardless of whether its edges are already present. The average 2-degree is then \( \langle k^{(2)} \rangle = \frac{(N - 1)(N - 2)}{2} p^{(2)} \). Each 2-simplex contributes, on average, \( 2(1 - p^{(1)}) \) additional links to the 1-degree of its nodes, depending on whether the corresponding edges (1-faces) were already present. As a result, the expected total 1-degree becomes:$\langle k^{(1)} \rangle_{\text{total}} \approx (N - 1)p^{(1)} + 2 \langle k^{(2)} \rangle (1 - p^{(1)})$. As a consequence, to craft a RSC with target values of \( \langle k^{(1)} \rangle \) and \( \langle k^{(2)} \rangle \), the connection probabilities to be chosen are:
\begin{align}
    p^{(1)} &= \frac{\langle k^{(1)} \rangle - 2 \langle k^{(2)} \rangle}{(N - 1) - 2 \langle k^{(2)} \rangle}, \label{eq:p1}\\
    p^{(2)} &= \frac{2 \langle k^{(2)} \rangle}{(N - 1)(N - 2)}. \label{eq:p2}
\end{align}

\subsubsection{Hypergraphs with maximum inter-order hyperedge overlap} 

If $2\langle k^{(2)}\rangle>\langle k^{(1)}\rangle$, Eq. (\ref{eq:p1}) yields negative values, as there are no enough $1-$hyperedges to fill all the faces of the $2-$hyperedges. Therefore, in these situations, we craft hypergraphs that maintain the downward closure to the best extent, i.e., with maximum inter-order hyperedge overlap \cite{lamata2025hyperedge}. In order to do so, we first create the set of $2-$hyperedges according to Eq. (\ref{eq:p2}), and afterwards, we sample as $1-$hyperedges a fraction of the faces of the $2-$hyperedges, according to the probability
\begin{eqnarray}
    \bar{p}^{(1)}=\frac{k^{(1)}}{2k^{(2)}}.
\end{eqnarray}

\medskip

Note that, in order to ensure the connectedness of the structure, we remove the nodes which are disconnected. This is the reason why the structure utilized in Figs. 1 and 2 has $N=1899$ instead of the original $N=2000$.

\subsection{Analytical derivation of the transition to consensus}
Neutral consensus corresponds to the fixed point $x_i=0\;\forall i\in\mathcal{N}$. To determine its stability we perform a mean field analysis from Eqs. (\ref{eq:der_x})-(\ref{eq:weights}) by assuming that all agents hold similar opinions, i.e., \( x_j \approx x \). We therefore approximate each group average as \( \sum_{j \in \gamma} x_j/m \approx x \), and the weights become uniform across hyperedges. This reduces Eq. (\ref{eq:der_x}) to 
\begin{equation}
    \dot x = -x+\sum_{m=1}^M \lambda^{(m)} \tanh(x). 
    \label{eq:mf_equation_diff}
\end{equation}
In the stationary state, where $\dot x=0$, Eq. (\ref{eq:mf_equation_diff}) yields a scalar fixed-point equation:
\begin{equation}
    x = \sum_{m=1}^M \lambda^{(m)} \tanh(x). \label{eq:mf_equation}
\end{equation}
Since \( \tanh(0) = 0 \), the solution \( x = 0 \) always exists. To assess its stability, we linearize around \( x = 0 \), using \( \tanh(x) \approx x \), yielding
\begin{equation}
    x \approx \left( \sum_{m=1}^M \lambda^{(m)} \right) x.
\end{equation}
Hence, \( x = 0 \) is stable if
\begin{equation}
    \sum_{m=1}^M \lambda^{(m)} < 1, \label{eq:stability_condition}
\end{equation}
and unstable otherwise. This condition defines the critical surface in parameter space where the neutral consensus loses stability and non-zero solutions (corresponding to polarization or radical consensus) emerge.
\medskip

\subsection{Numerical integration details}
The system of $N$ coupled differential equations is integrated using an explicit fourth-order Runge–Kutta method with a time step $dt = 0.1$.
\medskip

\subsection{Stability analysis of a polarized configuration}
The system of equations:
We perform a linear stability analysis of a polarized configuration, in which $N_+$ agents take a positive opinion $ x_+$ and $N_-=N-N_+$ a negative opinion $x_-$. In the Supplementary Eqs. (S.4)-(S.8) we derive that, in the equilibrium, both opinions fulfill the system of equations:
\begin{widetext}
\begin{align}
    x_{+} &= \sum_{m=1}^M \lambda^{(m)} \sum_{b=0}^m \left\{ n_m(b) \tanh \left(\frac{(m-b)x_++bx_-}{m}\right) \frac{f(b)^{\beta}}{\sum_{b'=0}^m n_m(b') f(b')^{\beta}} \right\} \, , \label{eq:x_+}\\
    x_{-} &= \sum_{m=1}^M \lambda^{(m)} \sum_{b=0}^m \left\{ n_m(b) \tanh \left(\frac{(m-b)x_++bx_-}{m}\right) \frac{f(m-b)^{\beta}}{\sum_{b'=0}^m n_m(b') f(m-b')^{\beta}} \right\} \, , \label{eq:x_-}
\end{align}
\end{widetext}

\noindent where 
\begin{align}
    n_m(b) = {{N_+}\choose{m-b}} {{N_-}\choose{b}} \, 
\end{align}
counts the number of possible $m$-hyperedges containing \( m-b \) agents with opinion \( x_+ \) and \( b \) agents with opinion \( x_- \). These terms reflect the statistical weight of each group configuration in a polarized population. Moreover, the function
\begin{align}  
    f(b) = \frac{\epsilon^{(m)}}{b(x_+-x_-)+\epsilon^{(m)}} \, 
\end{align}
defines the homophilic influence weight associated with a group where there are $b$ agents with opinion $x_-$. The function decays as the internal disagreement in the group increases (i.e., as $b$ grows), capturing the idea that agents are less influenced by dissimilar peers. Note that the small constant \( \epsilon^{(m)} \) ensures regularization near zero opinion difference.
\medskip

The polarized configuration given by Eqs.~(\ref{eq:x_+})-(\ref{eq:x_-}) is stable under perturbation of one of the agents if the following condition applies (see Supplementary Eqs. (S.8)-(S.21) for the analytical derivation): 

\begin{widetext}
\begin{equation}
    1 < \beta\sum_{m=1}^M \lambda^{(m)} \sum_{b=0}^m \left\{ n_m(b) \tanh\left[\frac{(m-b)x_++bx_-}{m}\right] \frac{f(b)^\beta \left[ \sum_{b'=0}^m n_m(b')f(b')^{\beta-1}g(b')\right] - f(b)^{\beta-1}g(b) \left[\sum_{b'=0}^m n_m(b')f(b')^\beta\right]}{\left[\sum_{b'=0}^m n_m(b')f(b')^\beta\right]^2}\right\} \, , \label{eq:stab}
\end{equation}
\end{widetext}

\noindent where the function
\begin{equation}
    g(b) = \frac{mb(x_+-x_-)+2b\epsilon^{(m)}}{\left[b(x_+-x_-)+\epsilon^{(m)}\right]^2} \,
\end{equation}
appears in the derivative of the weighted influence term $f(b)$ and captures how sensitive the homophilic weight is to small perturbations in the agent’s opinion. It depends on the opinion gap \( x_+ - x_- \), the group composition \( b \), and the small constant \( \epsilon^{(m)} \) to avoid divergence.

The full stability condition in Eq.~\eqref{eq:stab} ensures that small perturbations around the polarized configuration decay over time. Intuitively, it expresses a balance between the strength of homophilic influence (parametrized by \( \beta \)), the intensity of interactions \( \lambda^{(m)} \), and the structure of group compositions. If this condition is violated, small perturbations grow and the polarized configuration becomes unstable.
\medskip

\begin{acknowledgments}
\section{Acknowledgements}
H.P.M, S.L.O, and J.G.G. acknowledge financial support from the Departamento de Industria e Innovaci\'on del Gobierno de Arag\'on y Fondo Social Europeo (FENOL group grant E36-23R) and from Ministerio de Ciencia e Innovaci\'on (grant PID2020-113582GB-I00). H.P.M and S.L.O. acknowledge financial support from Gobierno de Aragón through a doctoral fellowship. D.S.-P. acknowledges financial support through grants JDC2022-048339-I and PID2021-128005NB- C21 funded by MCIN/AEI/10.13039/501100011033 and the European Union “NextGenerationEU”/PRTR”.
\end{acknowledgments}

\bibliography{biblio}
\clearpage

\renewcommand{\figurename}{Supplementary Fig.}
\renewcommand{\tablename}{Supplementary Table}
\renewcommand{\theequation}{S.\arabic{equation}}

\setcounter{equation}{0}
\setcounter{figure}{0}

\onecolumngrid
\newpage
\section{Supplementary Material of {\em Social polarization promoted by sparse higher-order interactions}}
\medskip

\section{Influence of hyperedge overlap in the emergence of polarization}

In higher-order structures, the \textit{hyperedge overlap} encodes how different groups of different orders share coincident sets of nodes. More specifically, the \textit{inter-order hypererdge overlap} measures the overlap between hyperedges of different orders, and is particularly relevant, as it clearly discriminates between \textit{simplicial complexes} (which maximize the ovelap due to their \textit{downward closure} property,  having every subset of nodes within a group as a valid lower-order interaction) and \textit{random hypergraphs} (where the overlap tends to zero, as randomness erases structural correlations). 

Following Ref. \cite{lamata2025hyperedge}, we define the \textit{inter-order hyperedge overlap} \( I^{(m,n)} \) as the fraction of \( m \)-node subsets (or \textit{faces}) induced by \( n \)-hyperedges that are also explicitly present as \( m \)-hyperedges in the structure:
\begin{equation}
    I^{(m,n)} = \frac{|\mathcal{E}^{(m)} \cap F(\mathcal{E}^{(n)})|}{|\mathcal{F}(\mathcal{E}^{(n)})|},
\end{equation}
where \( \mathcal{E}^{(m)} \) is the set of all \( m \)-hyperedges, and \( \mathcal{F}(\mathcal{E}^{(n)}) \) is the set of \( m \)-node subsets generated by the \( n \)-hyperedges.

Here, we study the influence of hyperedge overlap in the emergence of polarization. As in the main text, we consider $M=2$ and therefore we use as control metric \( I\equiv I^{(1,2)} \), which quantifies the fraction of  1-hyperedges that are present among the pairwise projections of the 2-hyperedges. In Supplementary Fig.~\ref{fig:overlap} we appreciate how inter-order hyperedge overlap promotes polarization, as enforces the local neighborhood of nodes, what eases the coexistence of opposite opinions within the population. 

\begin{figure}[h]
    \centering
    \includegraphics[width=.9\linewidth]{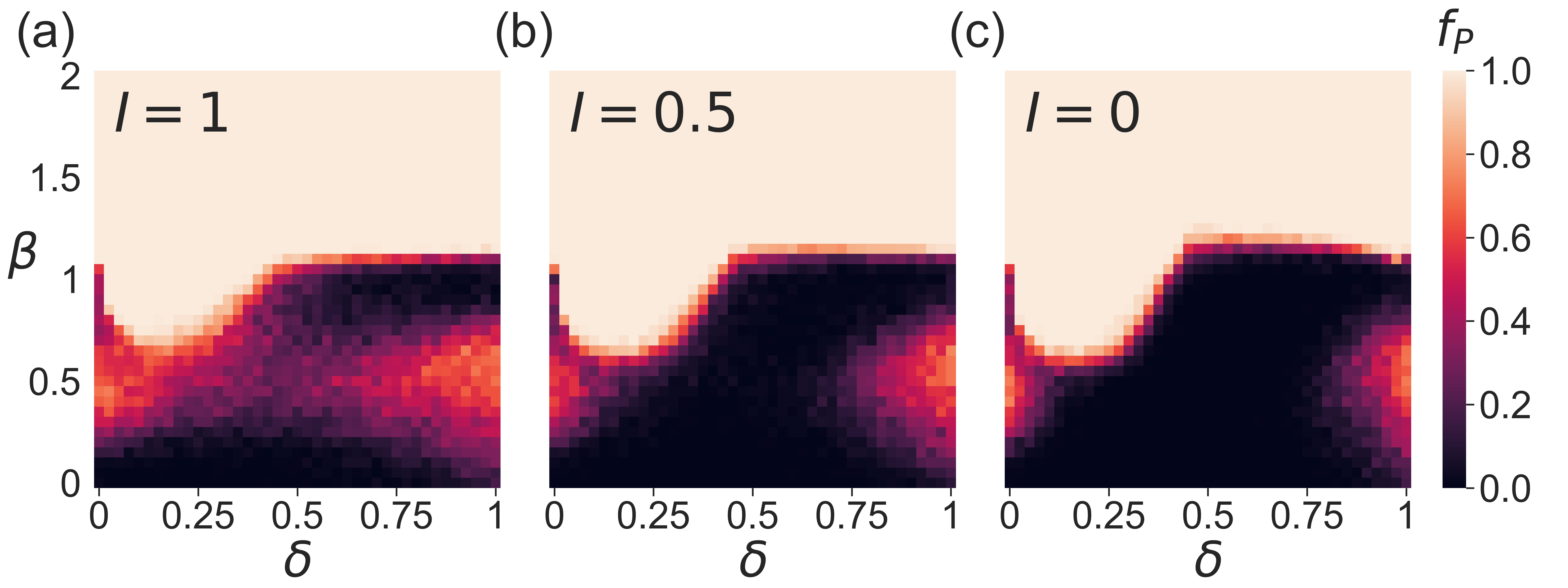}
    \caption{Fractions of polarized configurations $f_P$ obtained from 100 independent realizations for each combination of parameters $(\delta,\beta)$ (color code), and for different hyperedge overlap: (a) $I=1$, corresponding to the original simplicial complex. (b) $I=0.5$. (c) $I=0$. All networks are comprised of $N=1899$ nodes, with $\langle k^{(1)} \rangle=10$ and $\langle k^{(2)}\rangle=3$. Initial opinions are randomly chosen on the interval $[-20, 20]$.}
    \label{fig:overlap}
\end{figure}

\section{Stability analysis of a polarized configuration}

\subsection{Analytical derivation of the stationary configurations}

From Eq. (1)-(2) in the main text, we can directly obtain the equilibrium opinions $x_{\pm}$ of a fully connected hypergraph with $N_+$ agents of opinion $x_+$ and $N_-$ agents of opinion $x_-$. Considering that we have

\begin{equation}
    n_m(b) = {{N_+}\choose{m-b}} {{N_-}\choose{b}}
\end{equation}

\noindent hyperedges of order $m$ formed by $b$ neighbors of opinion $x_-$ and $m-b$ neighbors of opinion $x_+$, we can write:
\begin{align*}
    x_{+} &= \sum_{m=1}^M \lambda^{(m)} \left\{ \sum_{b=0}^m n_m(b) \tanh \left(\frac{(m-b)x_++bx_-}{m}\right) \frac{\left[b(x_+-x_-)+\epsilon^{(m)}\right]^{-\beta}}{\sum_{b'=0}^m n_m(b')\left[b'(x_+-x_-)+\epsilon^{(m)}\right]^{-\beta}} \right\} \, , \\
    x_{-} &= \sum_{m=1}^M \lambda^{(m)} \left\{ \sum_{b=0}^m n_m(b) \tanh \left(\frac{(m-b)x_++bx_-}{m}\right) \frac{\left[(m-b)(x_+-x_-)+\epsilon^{(m)}\right]^{-\beta}}{\sum_{b'=0}^m n_m(b')\left[(m-b')(x_+-x_-)+\epsilon^{(m)}\right]^{-\beta}} \right\} \, ,
\end{align*}

\noindent Now, dividing the numerator and denominator of the fractions by the smallest possible term, $({\epsilon^{(m)}})^{-\beta}$, we obtain:
\begin{align*}
    x_{+} &= \sum_{m=1}^M \lambda^{(m)} \left\{ \sum_{b=0}^m n_m(b) \tanh \left(\frac{(m-b)x_++bx_-}{m}\right) \frac{\left[\frac{\epsilon^{(m)}}{b(x_+-x_-)+\epsilon^{(m)}}\right]^{\beta}}{\sum_{b'=0}^m n_m(b')\left[\frac{\epsilon^{(m)}}{b'(x_+-x_-)+\epsilon^{(m)}}\right]^{\beta}} \right\} \, , \\
    x_{-} &= \sum_{m=1}^M \lambda^{(m)} \left\{ \sum_{b=0}^m n_m(b) \tanh \left(\frac{(m-b)x_++bx_-}{m}\right) \frac{\left[\frac{\epsilon^{(m)}}{(m-b)(x_+-x_-)+\epsilon^{(m)}}\right]^{\beta}}{\sum_{b'=0}^m n_m(b')\left[\frac{\epsilon^{(m)}}{(m-b')(x_+-x_-)+\epsilon^{(m)}}\right]^{\beta}} \right\} \, ,
\end{align*}

\noindent which, defining the function
\begin{equation}
    f(b) = \frac{\epsilon^{(m)}}{b(x_+-x_-)+\epsilon^{(m)}} \, ,
\end{equation}

\noindent results in:
\begin{align}
    x_{+} &= \sum_{m=1}^M \lambda^{(m)} \sum_{b=0}^m \left\{ n_m(b) \tanh \left(\frac{(m-b)x_++bx_-}{m}\right) \frac{f(b)^{\beta}}{\sum_{b'=0}^m n_m(b') f(b')^{\beta}} \right\} \, , \label{eq:x_+_supp}\\
    x_{-} &= \sum_{m=1}^M \lambda^{(m)} \sum_{b=0}^m \left\{ n_m(b) \tanh \left(\frac{(m-b)x_++bx_-}{m}\right) \frac{f(m-b)^{\beta}}{\sum_{b'=0}^m n_m(b') f(m-b')^{\beta}} \right\} \, ,\label{eq:x_-_supp}
\end{align}

\subsection{Analytical derivation of the stability condition}

\noindent Now, we introduce a perturbation in agent $i$, such that $x_i = x_+-\eta$. We denote the corresponding weight to a hyperedge of order $m$ with $b$ neighbors of opinion $x_-$ and $m-b$ of opinion $x_+$ as $w_{i}(m,b)$, whose expression results:
\begin{equation}
    w_{i}(m,b) = \frac{\left[(m-b)\eta + b(x_+-x_- -\eta)+\epsilon^{(m)}\right]^{-\beta}}{\sum_{b'=0}^m n_m(b')\left[(m-b')\eta + b'(x_+-x_- -\eta)+\epsilon^{(m)}\right]^{-\beta}} \, ,
\end{equation}

\noindent which, dividing by the smallest possible term, $(m\eta +\epsilon^{(m)})^{-\beta}$, results in:
\begin{equation*}
        w_{i}(m,b) = \frac{\left[\frac{m\eta +\epsilon^{(m)}}{(m-b)\eta + b(x_+-x_- -\eta)+\epsilon^{(m)}}\right]^\beta}{\sum_{b'=0}^mn_m(b')\left[\frac{m\eta +\epsilon^{(m)}}{(m-b')\eta + b'(x_+-x_- -\eta)+\epsilon^{(m)}}\right]^{\beta}} \, .
\end{equation*}

\noindent If $\eta$ is small enough, one can approximate:

\begin{align}
    \left[\frac{m\eta +\epsilon^{(m)}}{(m-b)\eta + b(x_+-x_- -\eta)+\epsilon^{(m)}}\right]^\beta &\simeq \left[\frac{\epsilon^{(m)}}{b(x_+-x_-) +\epsilon^{(m)}}\right]^\beta +  \nonumber\\
    &+ \beta \eta \left[\frac{\epsilon^{(m)}}{b(x_+-x_-)+\epsilon^{(m)}}\right]^{\beta-1} \frac{mb(x_--x_-)+2b\epsilon^{(m)}}{\left[b(x_+-x_-)+\epsilon^{(m)}\right]^2} \, .
\end{align}

\noindent Furthermore, defining the function:

\begin{equation}
    g(b) = \frac{mb(x_--x_-)+2b\epsilon^{(m)}}{\left[b(x_+-x_-)+\epsilon^{(m)}\right]^2} \, ,
\end{equation}

\noindent we finally have:

\begin{equation*}
    \left[\frac{m\eta +\epsilon^{(m)}}{(m-b)\eta + b(x_+-x_- -\eta)+\epsilon^{(m)}}\right]^\beta \simeq f(b)^\beta + \beta \eta f(b)^{\beta-1} g(b) \, ,
\end{equation*}

\noindent resulting in the following expression for the weights:
\begin{equation}
    w_i(m,b) = \frac{f(b)^\beta +  f(b)^{\beta-1}g(b)\beta\eta}{\sum_{b'=0}^m n_m(b') \left[f(b')^\beta +  f(b')^{\beta-1}g(b')\beta\eta\right]} \, ,
\end{equation}

\noindent which can be written as:
\begin{equation*}
    w_i(m,b) = \frac{f(b)^\beta +  f(b)^{\beta-1}g(b)\beta\eta}{\left(\sum_{b'=0}^m n_m(b') f(b')^\beta\right) +  \left(\sum_{b'=0}^m n_m(b')f(b')^{\beta-1}g(b')\right)\beta\eta} = \frac{A+B\eta}{C+D\eta}\, .
\end{equation*}

\noindent Furthermore, considering again small $\eta$, we can write:
\begin{equation}
    \frac{A+B\eta}{C+D\eta} \simeq \frac{A}{C} + \frac{BC-AD}{C^2}\eta \; ,
\end{equation}

\noindent and finally,
\begin{equation}
    w_i(m,b) \simeq \frac{f(b)^\beta}{\sum_{b'=0}^m n_m(b') f(b')^\beta} + \frac{f(b)^{\beta-1}g(b)\left[\sum_{b'=0}^m n_m(b')f(b')^\beta\right] - f(b)^\beta\left[\sum_{b'=0}^m n_m(b')f(b')^{\beta-1}g(b')\right]}{\left[\sum_{b'=0}^m n_m(b')f(b')^\beta\right]^2} \beta \eta \, .
    \label{Eq:app_weights}
\end{equation}

Taking all of this into account, the dynamic equation for agent $i$ results in:
\begin{equation}
    \dot{x}_i = -\dot{\eta} = - (x_+-\eta) + \sum_{m=1}^M \lambda^{(m)} \left[\sum_{b=0}^m n_m(b) w_i(m,b) \tanh\left(\frac{(m-b)x_++bx_-}{m}\right) \right] \;,
\end{equation}

\noindent where $w_i(m,b)$ follows Eq.~\ref{Eq:app_weights}. We notice that each weight has an independent part, and a linear term with $\eta$. If we group these separately, we reach the expression:
\begin{eqnarray}
    -\dot{\eta} &=& -(x_+-\eta) + \sum_{m=1}^N \lambda^{(m)} \left\{\sum_{b=0}^m n_m(b) \tanh\left(\frac{(m-b)x_++bx_-}{m}\right) \frac{f(b)^\beta}{\sum_{b'=0}^m n_m(b')f(b')^\beta} \right\}\\ &+& \beta \eta\sum_{m=1}^M\lambda^{(m)}\sum_{b=0}^m n_m(b) \tanh\left[\frac{(m-b)x_++bx_-}{m}\right]\frac{f(b)^{\beta-1}g(b)\left[\sum_{b'=0}^m n_m(b')f(b')^\beta\right] - f(b)^\beta\left[\sum_{b'=0}^m n_m(b')f(b')^{\beta-1}g(b')\right]}{\left[\sum_{b'=0}^m n_m(b')f(b')^\beta\right]^2}   \, \nonumber.
\end{eqnarray}

\noindent The independent term is equal to $x_+$, as can be seen in Eq.~\ref{eq:x_+_supp}. Therefore, we reach the expression:
\begin{equation}
    \frac{\dot{\eta}}{\eta} = -1 + \beta \sum_{m=1}^M \lambda^{(m)} \sum_{b=0}^m n_m(b) \tanh\left[\frac{(m-b)x_++bx_-}{m}\right] \frac{f(b)^\beta \left[ \sum_{b'=0}^m n_m(b')f(b')^{\beta-1}g(b')\right] - f(b)^{\beta-1}g(b) \left[\sum_{b'=0}^m n_m(b')f(b')^\beta\right]}{\left[\sum_{b'=0}^m n_m(b')f(b')^\beta\right]^2} \, .
\end{equation}

\noindent The stability condition corresponds to $\dot{\eta}>0$, and so, the fully polarized configuration will be unstable if:
\begin{equation}
    1 < \beta\sum_{m=1}^M \lambda^{(m)} \left\{\sum_{b=0}^m n_m(b) \tanh\left[\frac{(m-b)x_++bx_-}{m}\right] \frac{f(b)^\beta \left[ \sum_{b'=0}^m n_m(b')f(b')^{\beta-1}g(b')\right] - f(b)^{\beta-1}g(b) \left[\sum_{b'=0}^m n_m(b')f(b')^\beta\right]}{\left[\sum_{b'=0}^m n_m(b')f(b')^\beta\right]^2}\right\} \, , \label{eq:stab_supp}
\end{equation}

\noindent where, as a recall:
\begin{align}
    n_m(b) = {{N_+}\choose{m-b}} {{N_-}\choose{b}} \, , && f(b) = \frac{\epsilon^{(m)}}{b(x_+-x_-)+\epsilon^{(m)}} \, , && g(b) = \frac{mb(x_+-x_-)+2b\epsilon^{(m)}}{\left[b(x_+-x_-)+\epsilon^{(m)}\right]^2} \, .
\end{align}

\end{document}